\documentstyle[eqsecnum,preprint,aps]{revtex}

\begin{document}
\draft
% preprint ON
\preprint{WU-AP/52/95, gr-qc/9512045}
\title{Dynamics of gravitating magnetic monopoles}
\author{Nobuyuki Sakai\thanks{Electronic address: sakai@cfi.waseda.ac.jp}}
\address{Department of Physics, Waseda University, Shinjuku-ku, Tokyo 169,
Japan}
% preprint ON
\date{Revised 9 March 1996}
\maketitle

\begin{abstract}

According to previous work on magnetic monopoles, static regular solutions
are nonexistent if the vacuum expectation value of the Higgs field $\eta$
is larger than a critical value $\eta_{{\rm cr}}$, which is  of the order
of the Planck mass. In order to understand the properties of monopoles for
$\eta>\eta_{{\rm cr}}$, we investigate their dynamics numerically. If
$\eta$ is large enough ($\gg\eta_{{\rm cr}}$), a monopole expands
exponentially and a wormhole structure appears around it, regardless of 
coupling constants and initial configuration. If $\eta$ is around
$\eta_{{\rm cr}}$, there are three types of solutions, depending on
coupling constants and initial configuration: a monopole either expands as
stated above, collapses into a black hole, or comes to take a stable
configuration.

\end{abstract}

\vskip 1cm
\begin{center}
To appear in {\it Physical Review D}
\end{center}
%PACS number(s): 04.70.Bw, 11.15.-q, 14.80.Hv, 98.80.Cq

%%%%%%%%%%%%%%%%%%%%%%%%%%%%%%%%%%%%%%%%%%%%%%%%%%%%%%%%%%%%%%%%%%%%%%%%%%%%%%%
\newpage
% preprint ON
\baselineskip = 17pt
\section{Introduction}

In recent years static and spherically symmetric solutions of the
Einstein-Yang-Mills-Higgs system have been intensively studied in the
literature \cite{breit,lee,tachi}. One purpose of such investigation has
been to understand the nature of black holes, especially in the context
of the no-hair conjecture; it was shown that non-trivial black holes are
stable and hence the monopole black hole could be one of the most
plausible counterexamples. The other interest has been in the 
properties of particle-like solutions; it was shown that such regular
monopoles exist only if the vacuum expectation value of the Higgs field
$\eta$ is less than a critical value $\eta_{{\rm cr}}$, which is of the
order of the Planck mass $m_{Pl}$. This result naturally gives rise to
the next question: what is the fate of monopoles for $\eta>\eta_{{\rm
cr}}$?

Because the only static solution for $\eta>\eta_{{\rm cr}}$ is the
Reissner-Nordstr\"{o}m black hole, we can expect that a monopole 
which is regular initially evolves into the Reissner-Nordstr\"{o}m black
hole. Even if this speculation is reasonable, we still do not know how
the black hole formation occurs. One could imagine two alternatives: a
monopole just  shrinks, or its core continues to expand inside the
black-hole horizon, just as a ``child universe" \cite{sato}. 

Linde and Vilenkin independently pointed out the latter possibility in 
the context of the ``topological inflation" model \cite{topo}. They
claimed monopoles as well as other topological defects expand
exponentially if $\eta>O(m_{{\rm Pl}})$. Their discussions for the
Einstein-Higgs system were verified by our numerical simulation in
\cite{sakai}: we found that domain walls and global monopoles inflate if
and only if $\eta\stackrel{>}{\sim}0.33m_{{\rm Pl}}$. The next question
on this monopole inflation is similarly what happens to magnetic
monopoles in the Einstein-Yang-Mills-Higgs system. Because we cannot
find an answer to the question only by analyzing  static solutions, we
investigate dynamic monopole solutions in this paper. 

The plan of this paper is as follows. In Sec. II, we derive the basic
equations and explain how we solve those dynamical equations numerically. In
Sec. III, we offer analytic discussions and numerical results. Sec. IV is
devoted to summary and discussions. In this paper we use the units 
$c=\hbar=1$.

\section{Basic Equations}

The $SO(3)$ Einstein-Yang-Mills-Higgs system is described by
\begin{equation}\label{action}
  S=\int d^4 x \sqrt{-g} \left[\frac{m_{{\rm Pl}}^{~2}}{16\pi}{\cal R}
     -\frac14(F^a_{\mu\nu})^2
     -\frac12(D_{\mu}\Phi^a)^2-V(\Phi)\right],
\end{equation} with \begin{equation}\label{pote}
V(\Phi)= {1\over 4}\lambda(\Phi^2-\eta^2)^2, ~~ 
\Phi\equiv\sqrt{\Phi^a\Phi^a},
\end{equation} \begin{equation}
F^a_{\mu\nu} \equiv \partial_{\mu}A^a_{\nu}-\partial_{\nu}A^a_{\mu}
  +e\epsilon^{abc}A^b_{\mu}A^c_{\nu},~~
D_{\mu}\Phi^a\equiv\nabla_{\mu}\Phi^a+e\epsilon^{abc} 
A^b_{\mu}\Phi^c,
\end{equation}
where $A^a_{\mu}$ and $F^a_{\mu\nu}$ are the $SU(2)$ Yang-Mills field 
potential and its field strength, respectively. $\Phi^a$ is the real
triplet Higgs  field
and $V(\Phi)$ is its potential. $\lambda$ and $e$ are the Higgs self
coupling constant and the gauge coupling constant, respectively.
$\nabla_{\mu}$ and $D_{\mu}$ are the spacetime covariant derivative 
and the totally covariant derivative, respectively. The variation of
(\ref{action}) with respect to $g_{\mu\nu}$, $\Phi^a$ and $A^a_i$ yield
the Einstein equations,
$$
G_{\mu\nu}\equiv{\cal R}_{\mu\nu}-\frac12g_{\mu\nu}{\cal R}=
{8\pi\over m_{{\rm Pl}}^{~2}}T_{\mu\nu},
$$
\begin{equation}\label{ein}
T_{\mu\nu}\equiv D_{\mu}\Phi^aD_{\nu}\Phi^a
-g_{\mu\nu}\left[\frac12(D_{\sigma}\Phi^a)^2+V(\Phi)\right]
+F^a_{\mu\lambda}F^{a\lambda}_{\nu}-\frac14g_{\mu\nu}(F^a_
{\lambda\sigma})^2,
\end{equation}
and the equations for the matter fields,
\begin{equation}\label{heq}
D_{\mu}D^{\mu} \Phi^a=\frac{\partial V(\Phi)}{\partial\Phi^a},
\end{equation}
\begin{equation}\label{geq}
D_{\mu}F^{a\mu\nu}=-e\epsilon^{abc}\Phi^bD^{\nu}\Phi^c.
\end{equation}

We assume a spherically symmetric spacetime and adopt the coordinate 
system, 
\begin{equation}\label{metric}
ds^2=-dt^2+A^2(t,r)dr^2+B^2(t,r)r^2(d\theta^2+\sin^2\theta d\varphi^2).
\end{equation}
For the matter fields, we adapt the 't Hooft-Polyakov ansatz in such a 
way that
we can apply it to a time-dependent curved spacetime:
\begin{equation}\label{tp1}
\Phi^a=\Phi(t,r)\hat r^a,~~~
\hat r^a \equiv 
(\sin\theta\cos\varphi,\sin\theta\sin\varphi,\cos\theta).
\end{equation}
\begin{equation}\label{tp2}
A^a_{\mu}={\partial(\sqrt{g_{\theta\theta}}\hat r^b)
\over\partial x^{\mu}} \epsilon^{abc}\hat r^c{1-w(t,r)\over
e\sqrt{g_{\theta\theta}}}.
\end{equation}
With the metric (\ref{metric}) and the 't Hooft-Polyakov ansatz (\ref
{tp1}) and 
(\ref{tp2}), we write down the field equations (\ref{ein})-(\ref{geq})
as
\begin{eqnarray}\label{hc}
-G^0_0&\equiv&K^2_2(2K-3K^2_2)-{2B''\over A^2B}
-{B'^2\over A^2B^2} +{2A'B'\over A^3B}-{6B'\over A^2Br}
+{2A'\over A^3r} -{1\over A^2r^2}+{1\over B^2r^2} \nonumber\\
&=& {8\pi\over m_{{\rm Pl}}^{~2}}
\left[{\dot\Phi^2\over2}+{\Phi'^2\over2A^2}+\Bigl({w\Phi\over 
Br}\Bigr)^2+V
+{1\over(eBr)^2}\left\{\dot w^2+{w'^2\over A^2}
+\frac12\Bigl({w^2-1\over Br}\Bigr)\right\} \right], \\
%%%%%%%%
\frac12G_{01}&\equiv& {K^2_2}'+\Bigl({B'\over B}+{1\over r}\Bigr)(3K^
2_2-K)
={4\pi\over m_{{\rm Pl}}^{~2}}\left[\dot\Phi\Phi'+ {2\dot ww'\over
(eBr)^2}\right],
\label{mc}\end{eqnarray}
%%%%%%%%
$ \displaystyle \frac12(G^1_1+G^2_2+G^3_3-G^0_0) \equiv
\dot K-(K^1_1)^2-2(K^2_2)^2 $
\begin{equation}\label{dotk}
= {8\pi\over m_{{\rm Pl}}^{~2}} \left[\dot\Phi^2-V
+{1\over(eBr)^2}\left\{\dot w^2+{w'^2\over A^2}
+\frac12\Bigl({w^2-1\over Br}\Bigr)\right\} \right],
\end{equation}
%%%%%%%%
\begin{equation}\label{ddotphi}
\ddot\Phi-K\dot\Phi-{\Phi''\over A^2}-\left(-{A'\over A}+{2B'\over B}
+{2\over r}\right){\Phi'\over A^2} +{2w^2\Phi\over B^2r^2}+{dV\over 
d\Phi}=0,
\end{equation}
\begin{equation}\label{ddotw}
\ddot w -K^1_1\dot w-{w''\over A^2}+{A'w'\over A^3}-{w(1-w^2)\over 
B^2r^2}
+e^2\Phi^2w = 0,
\end{equation}
where the overdot and the prime denote the partial derivative with respect
to $t$ and $r$, respectively. We have introduced the extrinsic curvature 
tensor of a $t=$ constant hypersurface, $K_{ij}$, whose components are
given by
\begin{equation}\label{kdef}
K^1_1=-{\dot A\over A},~~~ K^2_2~(=K^3_3)~=-{\dot B\over B},
\end{equation}
and we have denoted its trace by $K\equiv K^i_i$. 

As an initial configuration of the matter fields, we adopt
the functional form of the static solution in a flat spacetime with
$\lambda=0$:
\begin{eqnarray}\label{init}
&&\Phi(t=0,r)=\Phi_{{\rm flat}}\Bigl({r\over c_{\Phi}}\Bigr)\equiv
\eta\left[{1\over\tanh(e\eta r/c_{\Phi})} - {1\over e\eta
r/c_{\Phi}}\right],\nonumber\\
&&w(t=0,r) = w_{{\rm flat}}\Bigl({r\over c_w}\Bigr)\equiv{e\eta
r/c_w\over\sinh(e\eta r/c_w)},
\end{eqnarray}
where $c_{\Phi}$ and $c_w$ are the initial size parameters. The
configurations of $\Phi_{{\rm flat}}(r)$ and of $w_{{\rm flat}}(r)$ are
illustrated in Fig. 1. As to the time-derivative, we suppose
$\dot\Phi(t=0,r)=\dot w(t=0,r)=0$. In order to set up consistent initial
data, we have to solve the constraint equations (\ref{hc}) and (\ref{mc}).
At this point, there are four unknown variables, $A,~B,~K$ and $K^2_2$, in
the two constraint  equations; two of the variables are arbitrarily chosen.
One of the methods which is usually adopted is to assume $K=$ const and
$A=B$. In this system, however, the condition of $K=$ const $\ne0$ is not
appropriate because the far region is asymptotically flat. Further, in the
range where there exists no static solution, we cannot fix $K=0$ even
momentarily. As an alternative, thereby, we suppose $A(t=0,r)=B(t=0,r)=1$
and solve the constraint equations (\ref{hc}) and (\ref{mc}) to determine
$K(t=0,r)$ and $K^2_2(t=0,r)$. This treatment is suitable for this system
because we obtain \begin{equation}
-{K\over3}\approx-K^2_2\approx\sqrt{{8\pi\over3m_{{\rm Pl}}^2}
\biggl({\Phi'^2\over2}+{\Phi^2\over r^2}+V\biggr)}, \end{equation} which
approaches zero as $r$ increases; we can construct an  asymptotically
flat spacetime without iterative integration. We have also assumed
$K(t=0,r)<0$: every point in the spacetime is locally expanding. The
numerical boundary is fixed at $r=30/(e\eta)^{-1}$ or $60/(e\eta)^{-1}$.

In order to solve the dynamical equations, we use a finite difference 
method with 2000 to 10,000 meshes. Now we have six dynamical variables:
$A,~B,~K,~K^2_2,~\Phi$ and $w$. Equations (\ref{dotk})-(\ref{kdef})
provide the next time-step of $A,~B,~K,~\Phi$ and  $w$, respectively. At
each step, we integrate (\ref{mc}) in the $r$-direction  to
obtain $K^2_2$. In this way we have reduced spatial derivatives 
appearing in the equations, which may become seeds for numerical
instability. The  Hamiltonian
constraint equation (\ref{hc}) remains unsolved during the evolution 
and is used for checking numerical accuracy. We stop numerical
computation when  some errors exceed a few percent.

In order to understand the spacetime structure from the numerical 
data, it is useful to observe the signs of the expansion of a null
geodesic congruence. Nambu and Siino \cite{nambu} also utilized this
tool to study wormhole formation in a singlet scalar field system. For
the  metric
(\ref{metric}), the expansion $\Theta_{\pm}$ is written as
\begin{equation}
\Theta_{\pm} =
{k^{2}_{\pm;2}}+{k^{3}_{\pm;3}}=2\bigg[-K^{2}_{2}\pm{(Br)'\over 
ABr}\biggl],
\end{equation}
where $k^{\mu}_{\pm}=(-1,\pm A^{-1},0,0)$ is an outgoing $(+)$ or 
ingoing
$(-)$ null vector. We observe the signs of $\Theta_{\pm}$ at all points 
in the numerical spacetime. For later convenience, we define ``RI" as
the region
where both $\Theta^+$ and $\Theta^-$ are positive, and ``RII" as the
region where they are negative. We can interpret that the region around
RI is de Sitter-like and the region around RII is Schwarzschild-like. The
two-surface which bounds RI or RII is called an ``apparent horizon".
Later, we will use the term ``black hole horizon" to refer to any
boundary of RII. And, we will use the term ``cosmological horizon" in
the sense that no information beyond it reaches the center of a monopole;
only the innermost boundary of RI is called the cosmological horizon.

\section{Numerical Results}

Before we move on to numerical simulation, we offer a rough discussion
on the effect of the gauge fields on the gravitational field. In our
previous paper \cite{sakai}, we investigated the effect for static
monopole solutions and found that the gauge fields generate an attractive
force. This property is also described
in the time-dependent coordinate system (\ref{metric}) as follows. 

In a homogeneous and isotropic spacetime, the evolution of the scale 
factor $a(t)$ follows
\begin{equation}\label{frw}
{3\ddot a\over a}=-{4\pi\over m_{{\rm Pl}}^{~2}}(\rho+3p),
\end{equation}
where $\rho$ and $p$ are the energy density and the pressure of a 
matter, respectively. Equation (\ref{frw}) indicates that the sign of
$\rho+3p$ determines whether the acceleration of the cosmic expansion is 
positive or negative. We can extend this discussion to general
spacetimes: the sign of $\rho+\Sigma p_i\equiv-T^0_0+T^i_i$ determines
whether a local
region expands with positive acceleration or not. A corresponding
equation in the present system to (\ref{frw}) is (\ref{dotk}). At the
origin, (\ref{dotk}) reduces to
\begin{equation}\label{center1}
{3\ddot A\over A}\Big|_{r=0}=
-{4\pi\over m_{{\rm Pl}}^{~2}}(\rho+\Sigma  p_i)_{r=0}\equiv
-{4\pi\over m_{{\rm Pl}}^{~2}}\Bigl(-2V+{3w''^2\over e^2A^4}\Bigr)_{r=0}
\end{equation}
In the case of global monopoles, the second term of the right-hand side
disappears, and hence the central region is always locally de Sitter 
spacetime. If gauge fields exist, however, the local acceleration at the
center also depends on the second term. Although the exact value of
$\partial^2w/(A\partial r)^2$ cannot be determined without solving the 
full dynamical equations, we can estimate its order by use of the
static solution in a flat spacetime with $\lambda=0$. Assuming
$\partial^2w/ (A\partial r)^2=b~w_{{\rm flat}}''(r)$ with $b=O(1)$, we have
\begin{equation}\label{center2}
\rho+\Sigma p_i\big|_{r=0}={e^2\eta^4\over4} \Bigl(-{\lambda\over e^2}
+\frac23b\Bigr),
\end{equation}
which lets us understand how the local expansion of the spacetime in the
center depends on $\lambda/e^2$. We see that, if $\lambda/e^2\ll1$, the
monopole core is an attractive spacetime; while, if $\lambda/e^2\gg1$,
it is repulsive like de Sitter spacetime. Of course, if the initial
configuration is quite different from that of the static solutions,
i.e., $b\ne O(1)$, the above discussion is not true. The dynamics may
also depend on initial configuration. 

In what follows, by use of the method in Sec. II, we will numerically 
integrate the field equations (\ref{hc})-(\ref{ddotw}). To show the
results, we define $X$ as a proper distance in the radial direction:
$\displaystyle X\equiv \int_0^r Adr$. We also define the boundaries of a
monopole in two ways: $X_{\Phi}(t)=X$ at the position of $\Phi=\eta/2$
and $X_{w}(t)=X$ at the position of $w=1/2$. We normalize time and length
by the horizon scale defined as $H_0^{-1}\equiv(8\pi V(0)/3m_{{\rm
Pl}}^{~2})^{-\frac12}$.

First, we check our numerical code by solving the equations for the case 
of weak gravity. In Fig. 2 we set $\eta=0.1m_{{\rm Pl}}$ and 
$\lambda/e^2=0.1$, and give two initial configurations: $c_{\Phi}=c_w=1$
and $0.5$. We plot the trajectories of $X_{\Phi}(t)$. We find that the
fields behaves stably; this reasonable result indicates that our numerical
code works well.

From now on, we concentrate on the parameter range where no static 
solution exists. In Fig. 3 we set $\eta=0.4m_{{\rm Pl}}$ and
$\lambda/e^2=0.1$, and give two initial configurations: $c_{\Phi}=c_w=1$
in (a) and $c_{\Phi}=c_w=10$ in (b). In Fig. 3(a)(b) we plot the
trajectories of $X_{\Phi}(t)$ and $X_w(t)$ as well as apparent
horizons. The dynamics in these two cases contrast sharply: in (a) a
monopole shrinks and black-hole horizons appear, while in (b) a
cosmological horizon exists from the beginning and a monopole continues to
expand. We also draw the distributions of $\rho+\Sigma p_i$ in (c) and in
(d), which correspond to the results in (a) and in (b), respectively. In (c)
the values around the center become negative at the beginning, but they
bounce back to positive values, which confirms that the monopole core never
inflates. On the other hand, in (d) the values of $\rho+\Sigma p_i$
around the center remain negative from the beginning. This behavior
indicates that exponential expansion really occurs inside the monopole.
These two results tell us that monopoles for $\eta>\eta_{{\rm cr}}$ tend to
be dynamic, and their dynamics depends on the initial configuration,
contrary to the case of global monopoles.

As we will see soon, for larger $\eta$, monopoles are more likely to inflate
rather than shrink. We show an example for larger $\eta$ in Fig. 4; we
set $\eta=0.55m_{{\rm Pl}}$, $\lambda/e^2= 0.1$ and $c_{\Phi}=c_w=1$. In
Fig. 4(a) we plot the trajectories of $X_{\Phi}(t)$ and $X_w(t)$ as well as
apparent horizons. (Please also refer to Fig. 5, which is a schematic sketch
of the spacetime structure.) From the beginning there are two apparent
horizons, $S1$ and $S2$: $S1$ is the cosmological horizon. Later other two
apparent horizons, $S3$ and $S4$, appear, and then  $S2$ and $S4$ approach
each other. These surfaces turn out to be black-hole horizons, $S2'$ and
$S4'$, the moment they intersect. In Fig. 4(b) we draw the distributions of
$\rho+\Sigma p_i$. Contrary to the case of the contracting monopole in Fig.
3(a), the values around the center are initially positive, but they become
negative. This suggests that, if $\eta$ is large enough, a monopole begins
to expand exponentially even if its initial size is not so large.  We also
show in Fig. 4(c) the relation between the proper distance along the radial
direction and the circumference radius, which indicates a wormhole
structure really appears. Figure 4 lets us understand how the wormhole is
created. Because the expanding core is causally disconnected from the outer
region, such an isolated region is called a ``child universe".

One may think that if $\eta>\eta_{{\rm cr}}$, a monopole either expands
or collapses, as shown in Fig. 3 or 4. However, we find some cases where a
monopole neither expands nor collapses. An example of such solutions is
shown in Fig. 6 ($\eta=0.3m_{{\rm Pl}}$ and $\lambda/e^2=1$). Setting
$c_{\Phi}=c_w=1$, we show the evolution of $\Phi$ in (a) and that of $w$ in
(b), and the trajectories of $X_{\Phi}(t)$ and of $X_w(t)$ in (c). Although
some oscillations remain outside the monopole, the core of the monopole
approaches a stable configuration. We change the initial size in Fig. 6(d),
finding monopoles with any initial size behave stably. These results
indicate that there exist stationary solutions.

In order to see if such stable monopoles are really created in an
expanding universe, we consider a different type of initial configurations:
to give a small perturbation on the symmetric state (de Sitter spacetime).
Specifically, we adopt a form,
\begin{equation}\label{init2}
\Phi(t=0,r)=d{r\over c} \exp\biggl[-\Bigl({r\over c}\Bigr)^2\biggr],  ~~~
w(t=0,r) =1,
\end{equation}
where we fix $d=0.1$ and $c=20$ in our analysis. In this way we see how a
monopole configuration is formed from the nearly symmetric state.
We assume $\lambda/e^2=10$ and $\eta/m_{{\rm Pl}}=0.3$, as is the case is
Fig. 6. Figure 7 shows that, once a monopole configuration is formed,
it approaches a stable configuration instead of continuing to expand. This
result supports the existence of stationary solutions. The existence of such
solutions looks surprising, because one may expect that all solutions in
the parameter range where no static solution exists must be dynamical. The
existence of ``stationary'' solutions, however, does not contradict the
nonexistence of ``static'' solutions: the solutions in Figs. 6 and 7 cannot
be described with a static coordinate system because the size of the
monopole is greater than the cosmological horizon.

Finally we systematically survey the dynamics of monopoles for
$0.05\le\eta/m_{{\rm Pl}}\le0.55$ and $0.1\le\lambda/e^2\le10$ and summarize
the solutions in the  $\lambda/e^2$-$\eta/m_{{\rm Pl}}$ plane of Fig. 7. A
square denotes a stable solution, as is the case in Fig. 2 or 6. A cross
denotes the case where a monopole shrinks, as is the case in Fig. 3(a). A
circle denotes the case where a monopole inflates and the wormhole
structure appears, as is the case in Fig. 4. A dotted line indicates the
maximum values of $\eta/m_{{\rm Pl}}$ versus $\lambda/e^2$, depicted
approximately by use of Fig.6 in \cite{breit}. We vary $c_{\Phi}$ and $c_w$
from 1 to 10, and hence some parameter points are labeled as two symbols.
We interpret these results as follows. In the case of $\lambda/e^2>1$, a
monopole expands exponentially if $\eta\stackrel{>}{\sim}0.35m_{Pl}$; this
critical value has little dependence on $\lambda/e^2$ and initial
configuration, and almost agrees with that for global monopoles
\cite{sakai}. This agreement is quite reasonable because the effect of the
gauge fields is smaller as $\lambda/e^2$ is larger. Below the critical
value, a monopole tends to take a stable configuration even in the theories
where static solutions are nonexistent. In the case of $\lambda/e^2<1$, the
dynamics also depend on $\lambda/e^2$ and initial configuration. In some
cases the effect of the gauge fields becomes dominant and a monopole
shrinks and becomes a black hole. These results are consistent with our
analytic discussions at the beginning of this section.

\section{Summary and Discussions}

We have studied the dynamics of magnetic monopoles numerically. Our
main purpose has been to understand the behavior of monopoles in the
case where static solutions are nonexistent.

If $\eta$ is large enough ($\gg\eta_{{\rm cr}}$), a monopole inflates and a
wormhole structure appears around it. We have shown how the wormhole
connected with a child universe is created. We should emphasize that a
child universe can be generated without fine-tuned initial conditions in
this model, contrary to the case of a trapped false vacuum bubble
\cite{sato}. In the case of $\lambda/e^2>1$, the condition of inflation is
$\eta\stackrel{>}{\sim}0.35m_{Pl}$, which has little dependence on
$\lambda/e^2$ and initial configuration. Below the critical value, a monopole tends to
take a stable configuration even in the theories where static solutions
are nonexistent. This is true for any initial configuration, which indicates
the existence of stationary solutions. While, in the case of $\lambda/e^2<1$,
the dynamics also depends on $\lambda/e^2$ and initial configuration. In
some cases the effect of the gauge fields becomes dominant and a monopole
collapses into a black hole.

We should notice that the condition of inflation was also estimated
analytically in a simplified model by Tachizawa et al.\cite{tachi} They
discussed the global structure of a spacetime by regarding the inside 
the monopole core as de Sitter spacetime and the outside as
Reissner-Nordstr\"{o}m spacetime. They showed that the surface of the 
monopole core exceeds a cosmological horizon if $\eta>m_{{\rm Pl}}/
\sqrt{3\pi}\approx 0.33m_{{\rm Pl}}$. This condition almost agrees with the
condition of inflation for most cases in our analysis; this agreement suggests
that our numerical results are reasonable as well as that their simplified
model is a good approximation in most cases. When the effect of the gauge
fields is dominant to that of the Higgs field, however, the spacetime is not de
Sitter-like, and then the validity of the simplified model is lost.

Our results as a whole support the discussions of Linde and Vilenkin
\cite{topo}, who pointed out the possibility of monopole inflation.
Actually, we have found that inflation happens in most cases of
$\eta>\eta_{{\rm cr}}$. Further, if the initial size of a monopole is large
enough, the effect of the gauge fields is not important, as Linde mentioned.
What we have clarified more about this subject is there are some cases where
static solutions are nonexistent but monopoles do not continue to expand, as the
results in Figs. 6 and 7. Although we did not state this fact in our
previous paper \cite{sakai}, it is also true for global monopoles.

% revtex
\acknowledgements

The author would like to thank J. Koga, A. Linde, K. Maeda, D. Maison, T.
Tachizawa, T. Torii, and A. Vilenkin for useful discussions. Thanks are also
due to P. Haines and W. Rozycki for reading the  manuscript. This work was
supported partially by the Grant-in-Aid for Scientific  Research Fund of the
Ministry of Education, Science and Culture (No.07740226) and by a
Waseda University Grant for Special Research Projects.

%%%%%%%%%%%%%%%%%%%%%%%%%%%%%%%%%%%%%%%%%%%%%%%%%%%%%%%%%%%%%%%%%%%%%%%%
%\newpage

%\newpage
\vskip 1cm
\baselineskip = 17pt
\noindent
{\bf Figure Captions}
\vspace{.3 cm}

\noindent
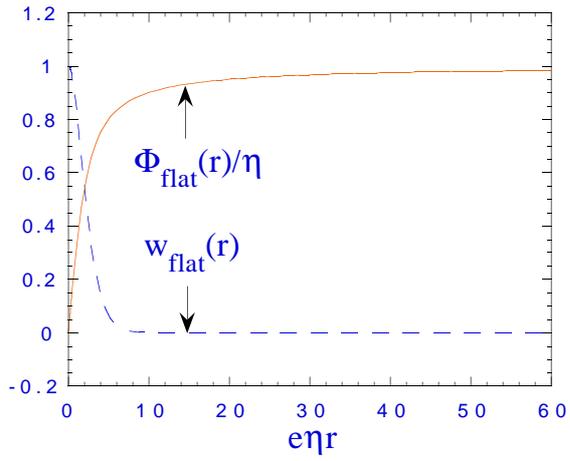
{\bf FIG.1}. Configurations of $\Phi_{{\rm flat}}(r)$ and of $w_{{\rm
flat}}(r)$ in (\ref{init}), which are adopted as initial conditions for
time-evolutions. 

\vskip .4cm\noindent
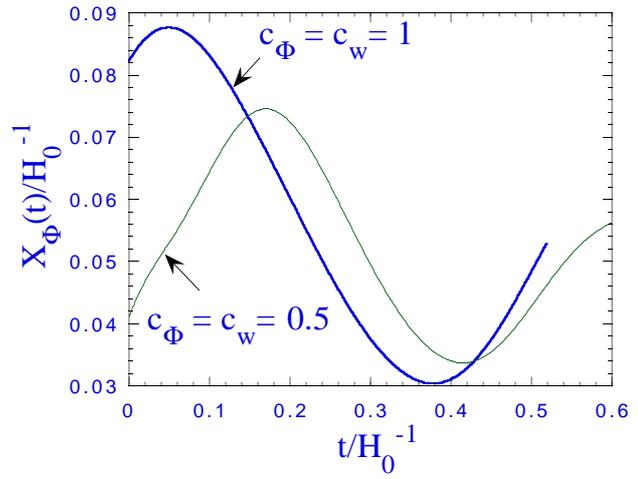
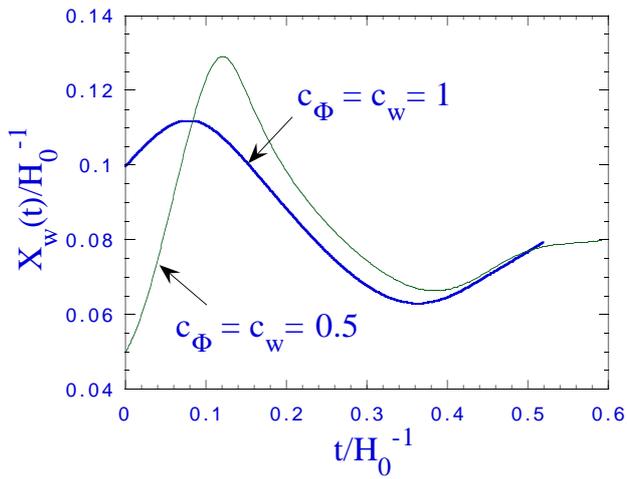
{\bf FIG.2}. Dynamics of a monopole for the case  of weak gravity. We
set $\eta=0.1m_{{\rm Pl}}$ and  $\lambda/e^2=0.1$, and give two initial
configurations: $c_{\Phi}=c_w=1$ and $0.5$. We plot the trajectories
of $X_{\Phi}(t)$ in (a) and of $X_w(t)$ in (b). The fields behave stably;
these reasonable results indicates that our numerical code works well.

\vskip .4cm\noindent
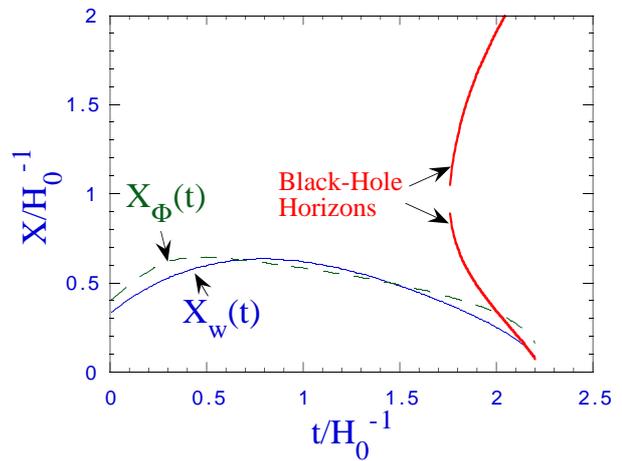
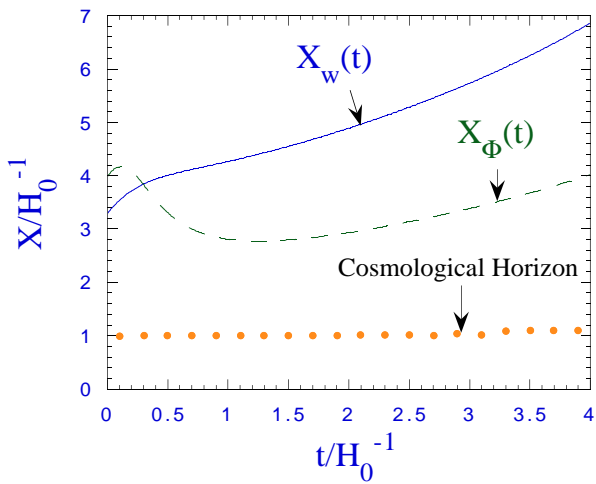
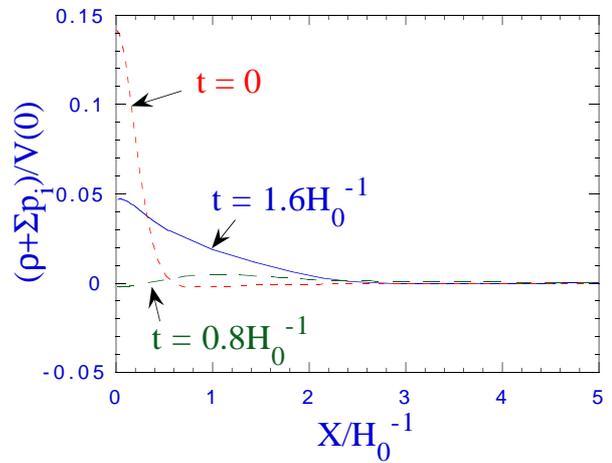
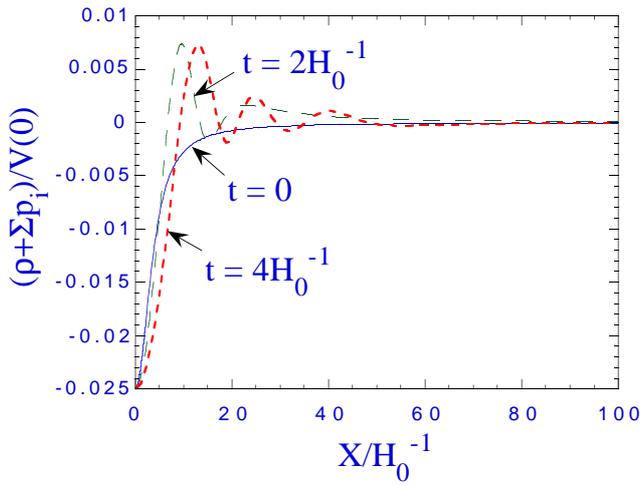
{\bf FIG. 3}. Dynamics of a monopole for $\eta=0.4m_{{\rm Pl}}$ and
$\lambda/e^2=0.1$. We assume two initial configurations: $c_{\Phi}=c_w=1$ in
(a) and $c_{\Phi}=c_w=10$ in (b). In (a) and (b) we plot the trajectories
of $X_{\Phi}(t)$ and $X_w(t)$ as well as apparent horizons. In (a) a
monopole shrinks and the black-hole horizons appear, while in (b) a
cosmological horizon exists from the beginning and a monopole continues to
expand. We also draw the distributions of $\rho+\Sigma p_i$ in (c) and in
(d), which correspond the results in (a) and in (b), respectively. In (c)
the values around the center get negative at the beginning, but they bounce
back to positive values, which confirms that the monopole core never
inflates. On the other hand, in (d) the values of $\rho+\Sigma p_i$ around
the center remain negative from the beginning. This behavior indicates that
exponential expansion really occurs inside the monopole.

\vskip .4cm\noindent
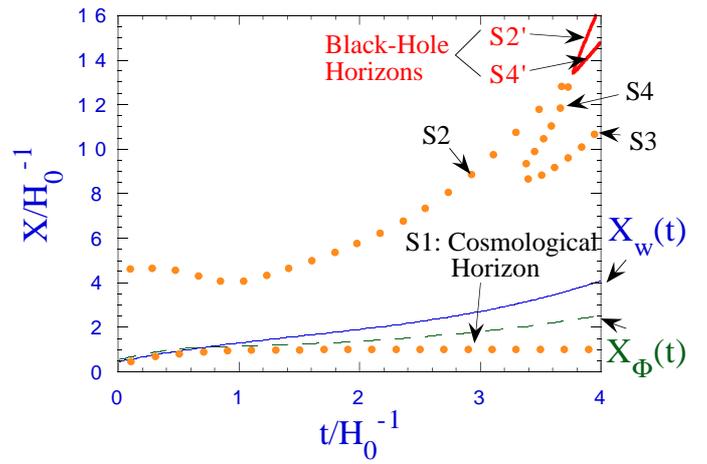
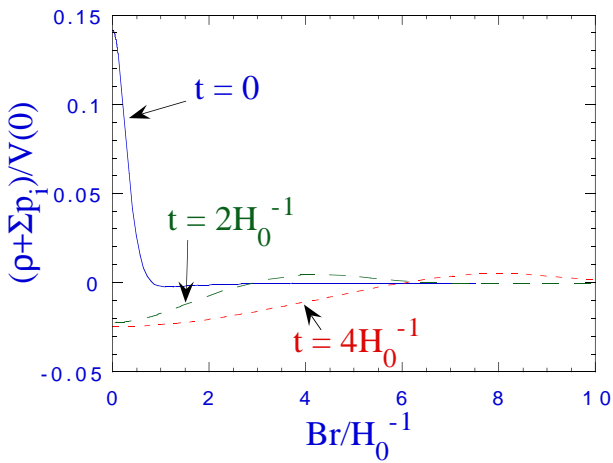
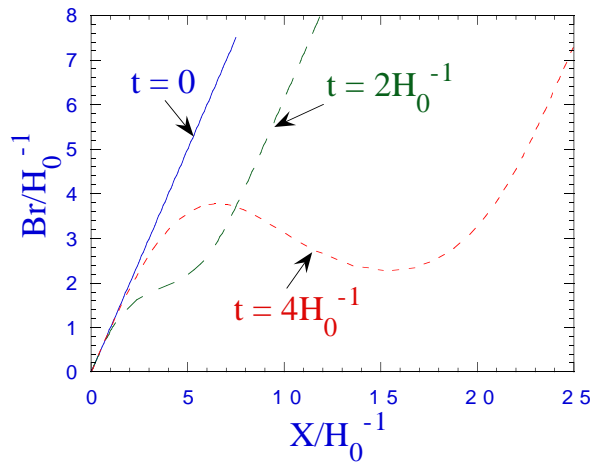
{\bf FIG. 4}. Dynamics of a monopole for $=0.55m_{{\rm Pl}}$, $\lambda/e^2=
0.1$ and $c_{\Phi}=c_w=1$. In (a) we plot the trajectories of
$X_{\Phi}(t)$ and $X_w(t)$ as well as apparent horizons. (Please also refer
to Fig. 5, which is a schematic sketch of the spacetime structure.) From
the beginning there are two apparent horizons, $S1$ and $S2$: $S1$ is the
cosmological horizon. Later other two apparent horizons, $S3$ and $S4$,
appear, and then $S2$ and $S4$ approach each other. These surfaces turn 
out to be black-hole horizons, $S2'$ and $S4'$, the moment they intersect.
In (b) we draw the distributions of $\rho+\Sigma p_i$. The values around the
center are initially positive, but they become negative. This suggests
that, if $\eta$ is large enough, a monopole begins to expand exponentially
even if its initial size is not so large. We also show in (c) the relation
between the proper distance along the radial direction and the
circumference radius, which indicates a wormhole structure really appears. 

\vskip .4cm\noindent
{\bf FIG. 5}. Schematic sketches of the spacetime structure. These 
figures are not generated from the numerical data, but they are based on
the results presented in Fig. 4. 

\vskip .4cm\noindent
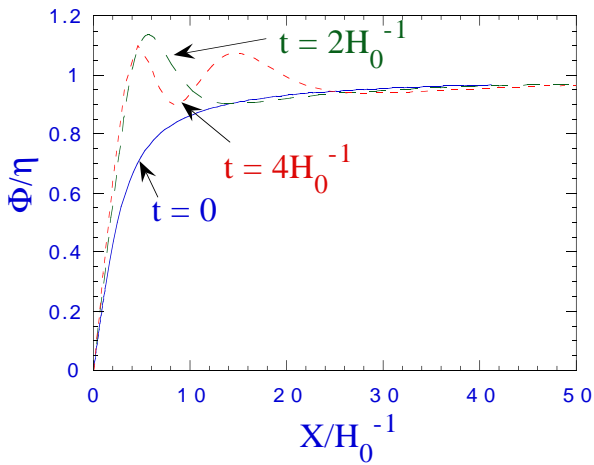
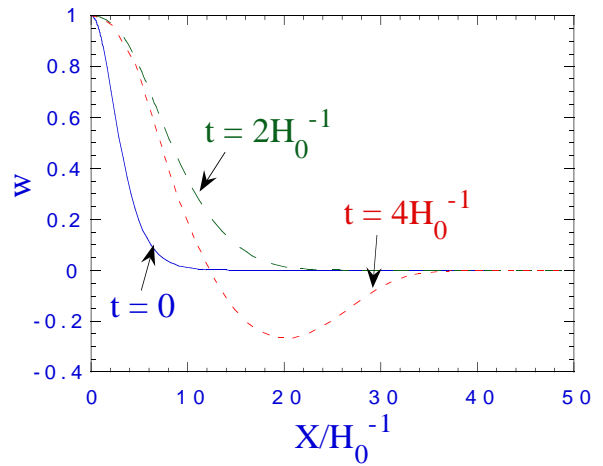
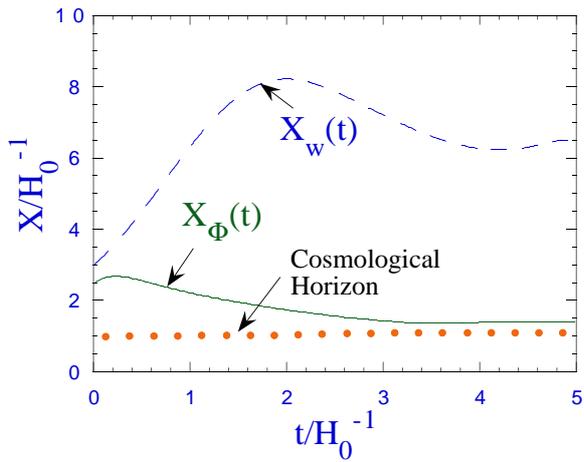
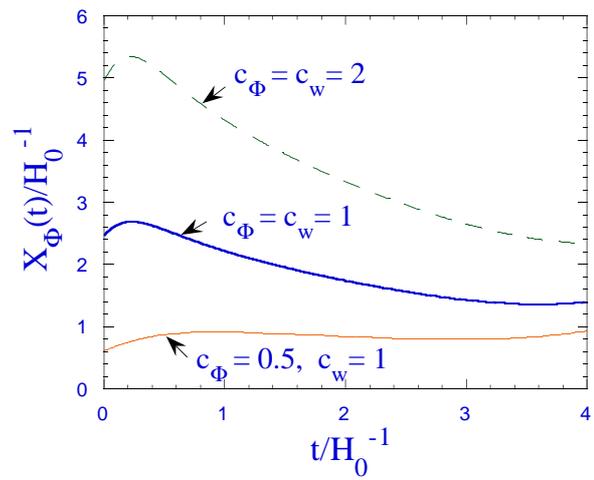
{\bf FIG. 6}. Dynamics of a monopole for $\eta=0.3m_{{\rm Pl}}$ and
$\lambda/e^2=1$. Setting $c_{\Phi}=c_w=1$, we show the evolution of $\Phi$ in
(a) and that of $w$ in (b), and the trajectories of $X_{\Phi}(t)$ and
of $X_w(t)$ in (c).  Although some oscillations remain outside the monopole,
the core of the monopole approaches a stable configuration. We change the
initial size in (d), finding monopoles with any initial size behave stably.
These results indicate that there exist stationary solutions.

\vskip .4cm\noindent
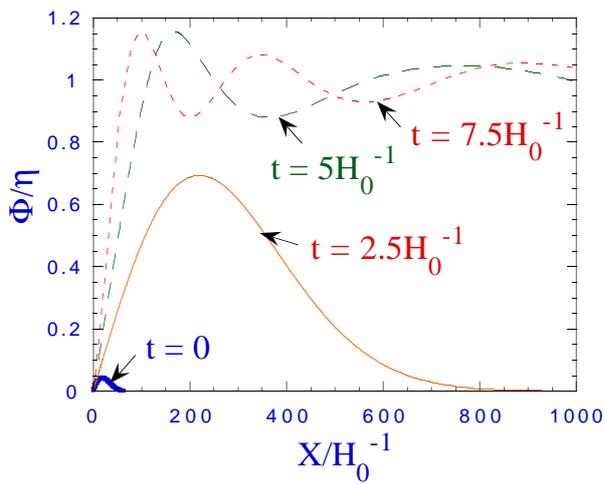
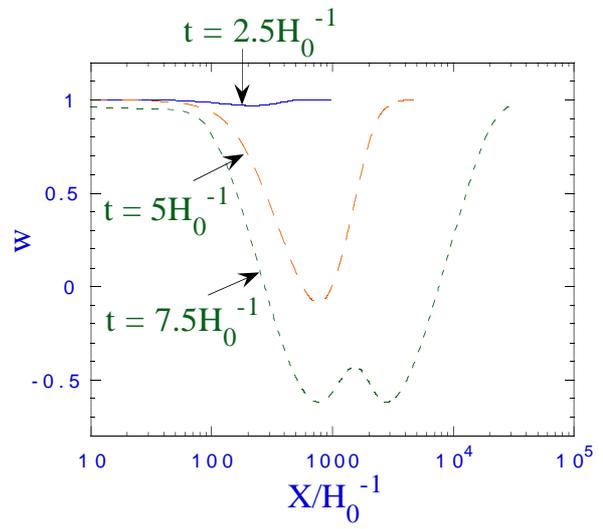
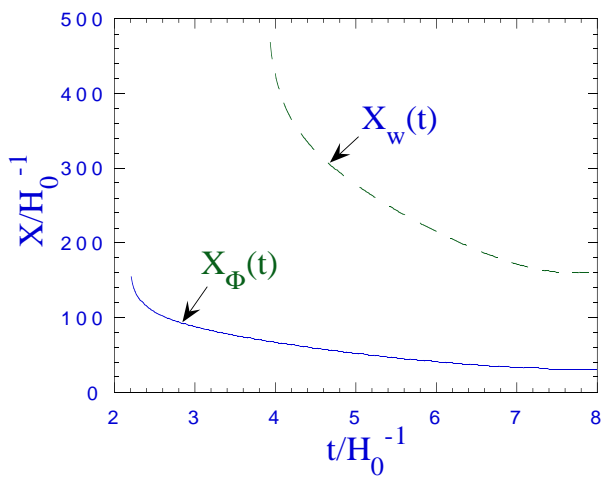
{\bf FIG. 7}. Formation of a monopole from the perturbed symmetric state
(de Sitter spacetime). An initial configuration is given by
(\ref{init2}). We draw the configuration of $\Phi$ in (a) and that of $w$ in
(b), and the trajectories of $X_{\Phi}(t)$ and $X_w(t)$ in (c). Once a
monopole configuration is formed, the monopole approaches a stable
configuration instead of continuing to expand. This result supports the
existence of stationary solutions. 

\vskip .4cm\noindent
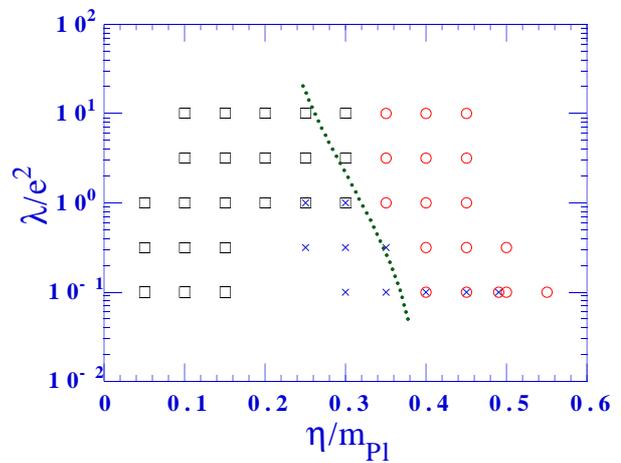
{\bf FIG. 8}. Parameter plane of $\eta/m_{{\rm Pl}}$ and $\lambda/e^2$ 
in which we summarize our numerical results. A square
$({\kern1pt\vbox{\hrule height 1.2pt\hbox{\vrule width1.2pt\hskip 3pt
\vbox{\vskip 6pt}\hskip 3pt\vrule width 0.6pt}\hrule height 0.6pt}\kern1pt}
)$ denotes a stable solution, as is the case in Fig. 2 or 6. A cross
$(\times)$ denotes the case where a monopole shrinks, as is the case in
Fig. 3(a). A circle $(\bigcirc)$ denotes the case where a monopole inflates
and the wormhole structure appears, as is the case in Fig. 3(c) or 4. A
dotted line indicates the maximum values of $\eta/m_{{\rm Pl}}$  versus
$\lambda/e^2$, depicted approximately by use of Fig.6 in \cite{breit}.
We vary $c_{\Phi}$ and $c_w$ from 1 to 10, and hence some parameter
points are labeled as two symbols. 

\end{document}